\documentclass[a4paper,11pt]{article}
\usepackage{pos}
\usepackage{dcolumn}   
\usepackage{bm}        
\usepackage{float}

\usepackage[]{caption}
\title{Reheating Effects on Charged Lepton Yukawa Equilibration and Leptogenesis}

\author*[a]{Rishav Roshan}

\affiliation[a]{School of Physics and Astronomy, University of Southampton, Southampton SO17 1BJ, United Kingdom}

\emailAdd{r.roshan@soton.ac.uk}

\abstract{We show that accounting for a non-instantaneous reheating phase after inflation can significantly modify the charged lepton Yukawa equilibration temperature in the early Universe. This finding calls for revisiting the role of lepton flavors in leptogenesis models where right-handed neutrinos are produced and decay during the extended reheating period. Our analysis reveals that this effect can induce shifts in the flavor regime(s) of leptogenesis relative to standard scenario.}

\FullConference{Proceedings of the Corfu Summer Institute 2025 "School and Workshops on Elementary Particle Physics and Gravity" (CORFU2025)\\
27 April - 28 September, 2025\\
Corfu, Greece\\}

\begin{document}
\maketitle

\section{Introduction}

The existence of non-zero neutrino masses~\cite{Fukuda:1998mi,Ahmad:2002jz,Ahn:2002up} and the baryon asymmetry of the Universe~\cite{Planck:2018vyg} (BAU) represent major challenges unaddressed by the Standard Model (SM), indicating physics beyond the SM is required. The light neutrino mass problem can be elegantly resolved by introducing three heavy SM singlet right-handed neutrinos (RHNs) with Yukawa interactions to SM Higgs and lepton doublets via the `seesaw' mechanism~\cite{Minkowski:1977sc,Yanagida:1979as,Yanagida:1979gs,GellMann:1980vs,Mohapatra:1979ia,Schechter:1980gr,Schechter:1981cv,Datta:2021elq}. This framework also explains the BAU through leptogenesis \cite{Fukugita:1986hr,Luty:1992un,Pilaftsis:1997jf}, where CP-violating out-of-equilibrium RHN decays generate a lepton asymmetry that is partially converted to baryon asymmetry via $(B+L)$ violating sphaleron interactions at electroweak temperatures $T_{\rm EW} \gtrsim$ 100 GeV.

In the well-studied `$thermal$' leptogenesis, once the Universe enters the radiation-dominated era at reheating temperature $T_{\text{RH}}$, RHNs are thermally produced through inverse decays and 2-2 scattering. For hierarchical RHN masses, the lightest ($N_1$ with mass $M_1$) produces lepton asymmetry via out-of-equilibrium decay to SM leptons ($l_{L_{\alpha}}$) and Higgs ($H$) at $ T \lesssim M_1$, with $T_{\text{RH}} > M_1$. The RHN abundance ($Y_{N_1}$) and lepton asymmetry in flavor $\alpha$ ($\Delta_{L_{\alpha}}$) are governed by Boltzmann equations incorporating all lepton-number violating processes.

When decay of RHN occurs above $5\times10^{11}$ GeV, where charged lepton Yukawa interactions remain out of equilibrium, the unflavored approximation applies \cite{Fukugita:1986hr,Luty:1992un,Pilaftsis:1997jf}. Below this temperature, the charged yukawa interaction of right-handed tau leptons equilibrate and flavor effects ~\cite{Barbieri:1999ma,Nardi:2005hs,Nardi:2006fx,Abada:2006fw,Abada:2006ea,Blanchet:2006ch,Blanchet:2006be,Dev:2017trv,Datta:2021gyi} become crucial. Equilibrated tau Yukawa interactions destroy tau-carried lepton asymmetry from RHN decays, with muon and electron Yukawa equilibration occurring below $10^{9}$ GeV and $5\times10^{4}$ GeV respectively.

It is well known that a lower bound on RHN mass exists as $M_1 \gtrsim 10^9$ GeV (the Davidson-Ibarra bound \cite{Davidson:2002qv}) to achieve the observed baryon asymmetry via leptogenesis, requiring $T_{\text{RH}} > M_1$ for standard {\it thermal} leptogenesis. While such high $T_{\text{RH}}$ is possible, there is no concrete evidence supporting it, and it could be as low as few MeV~\cite{Giudice:2000ex,Kawasaki:2000en,Martin:2010kz,Dai:2014jja}. This motivates exploring leptogenesis with $T_{\text{RH}} < M_1$. One option is {\it non-thermal} leptogenesis \cite{Lazarides:1991wu,Murayama:1992ua,Kolb:1996jt,Giudice:1999fb,Asaka:1999yd,Asaka:1999jb,Hamaguchi:2001gw,Jeannerot:2001qu,Fujii:2002jw,Giudice:2003jh,Pascoli:2003rq,Asaka:2002zu,Panotopoulos:2006wj,HahnWoernle:2008pq,Hamada:2015xva,Borah:2020wyc,Samanta:2020gdw,Barman:2021tgt,Azatov:2021irb,Barman:2021ost,Barman:2022gjo,Lazarides:2022spe,Lazarides:2022ezc,Ghoshal:2022fud,Ghoshal:2022kqp}. Another involves non-instantaneous reheating extending from $T_{\text{Max}}$ to $T_{\text{RH}}$. Reheating can indeed be a gradual process \cite{Chung:1998rq,Giudice:2000ex,Mukaida:2015ria,Harigaya:2019tzu,Garcia:2020eof,Haque:2020zco} where a maximum post-inflationary temperature $T_\text{Max}$ precedes the radiation-dominated era at $T_{\text{RH}}$ with $T_{\text{Max}} > T_{\text{RH}}$.

This proceeding is based on our work \cite{Datta:2022jic}, here we show that leptogenesis remains viable when $T_{\text{Max}} > M_1 > T_{\text{RH}}$. A more general picture allowing an additional interaction between the inflaton and the RHN fields on top of the effective coupling between the inflaton and SM fermion fields, can be found in \cite{Datta:2023pav}. Importantly, we find that prolonged reheating modifies the equilibration temperature (ET) of charged lepton Yukawa interactions, enriching flavor leptogenesis during this extended period. The effective inflaton coupling to SM fermions determines the radiation component and thermal bath temperature, non-trivially affecting the expansion rate during reheating. Sufficiently large effective coupling accelerates expansion, delaying charged lepton Yukawa equilibration \footnote{A recent study \cite{Roshan:2025mwl} shows that early matter domination after reheating can similarly shift the charged lepton Yukawa equilibration temperature. }. When RHNs are thermally produced and decay out of equilibrium at $T \lesssim M_1$ during extended reheating, this delayed equilibration significantly shifts the flavor regimes of leptogenesis.

\section{Charged Lepton Equilibration and Flavored leptogenesis}
\label{sec2}

To determine whether charged lepton Yukawa interaction of flavor $\alpha$ ($Y_{\alpha} \overline{\ell}_{L_\alpha} H e_{R_\alpha}$ with $e_{R\alpha}$ as representative of right handed electron/muon/tau) is in thermal equilibrium at a given temperature, the interaction rate ($\Gamma_{\alpha}$) must exceed the Universe's expansion rate \cite{Nardi:2006fx}. In the standard radiation-dominated scenario, the thermally averaged interaction rates for SM Higgs doublet decay to left-handed lepton doublets and right-handed charged lepton singlets ($H\leftrightarrow \ell_{L_\alpha} e_{R_\alpha}$\cite{Campbell:1991at}) are estimated as \cite{Campbell:1991at,PhysRevLett.71.2372,Cline:1993bd}:

	\begin{align}
		\langle\Gamma_\alpha \rangle=\int \frac{d^3 p} {(2\pi)^3 2E_P}\int \frac{d^3 k}{(2\pi)^3 2E_k}\int \frac{d^3 k^\prime}{(2\pi)^3 2E_{k^\prime}}
		(2\pi)^4 \delta^{(4)}(p-k-k^{\prime})|\mathcal{M}|^2 \frac{f_p}{n_p},
		\label{eq:decayrate}
	\end{align}

\noindent where $p$ is the $4$-momentum of the Higgs while $k$ and $k^{\prime}$ are the $4$-momentum of lepton doublet and singlet right handed charged lepton respectively. The thermal distribution of Higgs $f_p$ and number density $n_p$  are taken as:
\begin{align}
	f_p =\frac{1}{e^{E_p/T}-1}\,,~
	n_p=\frac{\zeta(3) T^3}{\pi^2}\,.
\end{align}
The matrix amplitude squared $|\mathcal{M}|^2$ for such decay would be (assuming final state particles have negligible mass):
\begin{align}
	|\mathcal{M}|^2= 2 Y_{\alpha}^2 k.k^{\prime}= Y_{\alpha}^2 M_H^2\,,\quad \alpha=e,\mu,\tau.
\end{align}
Evaluation of the integrals in Eq.~\eqref{eq:decayrate} for $T \gg M_H$ yields \cite{Campbell:1991at}:
\begin{align}
	\langle\Gamma_\alpha \rangle=\frac{Y_{\alpha}^2 \pi}{192 \zeta(3)T} M_H^2.
    \label{eq-temp}
\end{align}
Considering the thermal mass of the Higgs to be \cite{Weldon:1982bn,Quiros:1999jp,Senaha:2020mop}: 
\begin{align}
	M_H =	M_H(T)\simeq \frac{T}{4}\sqrt{3 g^2+g^{\prime^2}+ 4y_t^2+8 \lambda},
\end{align}
where $g,g^{\prime}$ are SM gauge coupling constants and $y_t$, $\lambda$ are the top Yukawa and the Higgs quartic couplings respectively, the thermally averaged interaction rate for the decay processes will become \cite{Abada:2006ea} $\mathcal{O} (5 \times 10^{-3}) Y_{\alpha}^2 T$.

However, thermal corrections for the final state particles can also be important as they are \cite{Weldon:1982bn}: $m_{\ell_L} (T)= \frac{1}{4} \sqrt{\left(3 g^2 + g'^2\right)}T$ and $m_{e_R} (T) = \frac{1}{2} g' T$ for $\ell_L$ and $e_R$ respectively. 
Though the hierarchy $M_H (T) > m_{\ell_L} (T) > m_{e_R} (T)$ is always maintained for $T > T_{EW}$, a 
situation can be achieved at some high temperature where this Higgs decay channel may actually be closed with 
$M_{H} (T)$ being smaller than $m_{\ell_L} (T) + m_{e_R} (T)$. This happens due to the decrease of top Yukawa coupling $y_t$ with the increase in temperature~\cite{Bodeker:2019ajh}. 
\begin{figure}[t]
\centering
	\includegraphics[width=0.5\linewidth]{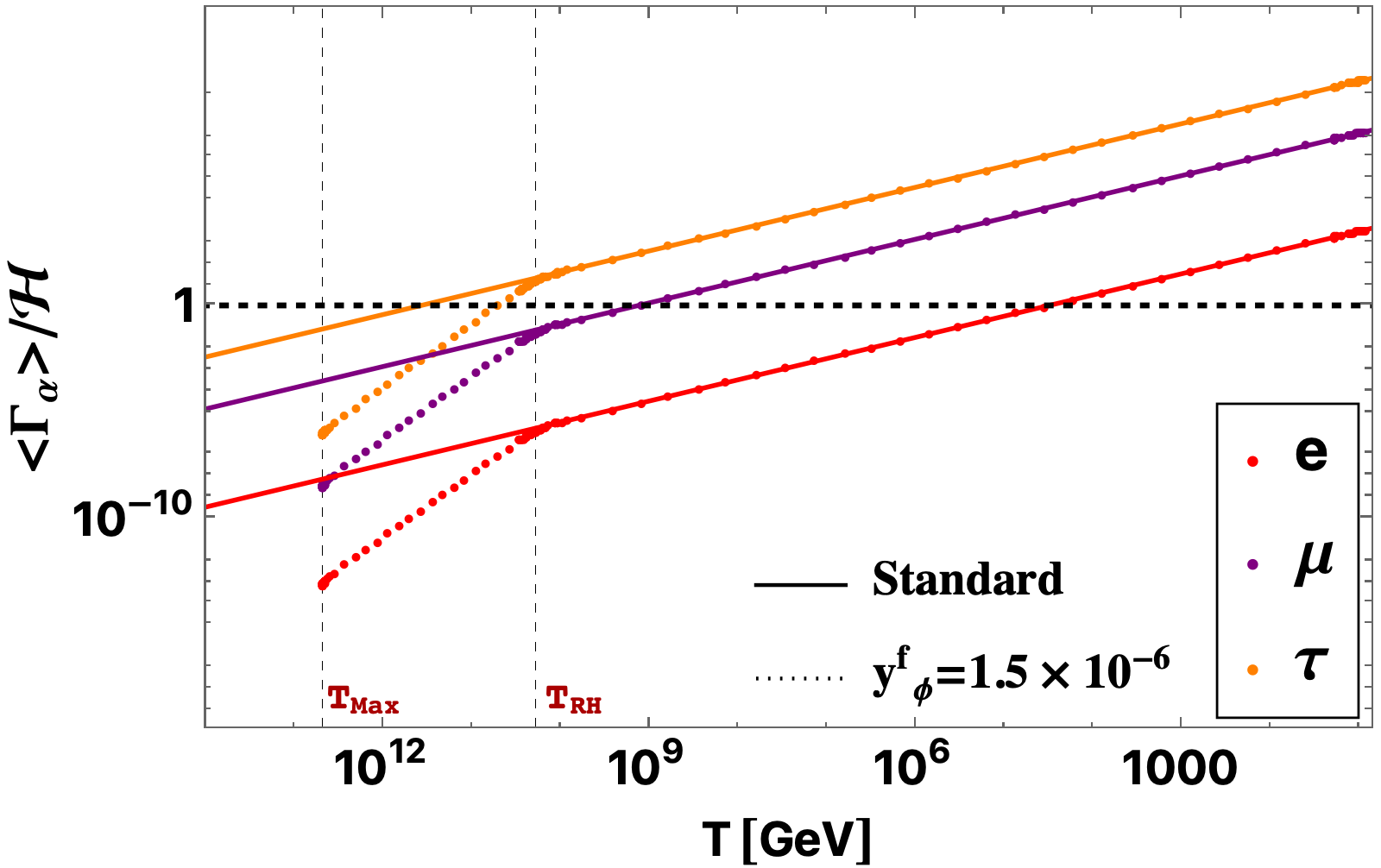}
	\caption{ Variation of $\langle\Gamma_\alpha \rangle/\mathcal{H}$ $w.r.t.$ $\rm{T}$ for standard (solid lines) and modified (dotted) scenarios. Horizontal dashed line denotes $\langle\Gamma_\alpha \rangle/H=1$, while the vertical dashed lines indicate the ETs of three charged Yukawa interactions.}\label{fig:1}
\end{figure}
Note that apart from the 1$\rightarrow$ 2 decay, the 2$\leftrightarrow$2 scatterings involving the specific Yukawa interaction $Y_{\alpha}$ (such as $X H^{\dagger} \rightarrow \bar{\ell}_L e_R, \ell_L H^{\dagger}  \rightarrow X e_R$ $etc.$ where $X = B, W$ gauge bosons) are also important at high temperature as their contributions to the interaction rate falls in the range: $(5.19 - 4.83) \times 10^{-3} ~Y^2_{\alpha} T$ \cite{Garbrecht:2013bia,Garbrecht:2014kda} which is found to be in the same ballpark of the naive decay estimate above. Hence, for the purpose of our study, we consider the interaction rate $\langle\Gamma_\alpha \rangle$ associated to the charged lepton Yukawa $Y_{\alpha}$ to be: $\langle\Gamma_\alpha \rangle\simeq 5 \times 10^{-3} Y_{\alpha}^2 T.$
Comparison between the obtained interaction rates $\langle\Gamma_\alpha \rangle$ with Hubble constant ($\mathcal{H}= 1.66 g_*^{1/2} T^2/M_{pl}$ in radiation dominated Universe, where $M_{pl}$ is the Planck mass.) will lead to the ET of right-handed charged lepton singlets. 
Solid lines in Fig.~\ref{fig:1} shows the variation of $\langle\Gamma_\alpha \rangle/{\mathcal{H}}$ with respect to temperature $T$ for different lepton flavors: $\alpha=e,\mu,\tau$. Note that $\tau$ Yukawa interaction becomes fast enough around $T = T^*_{0(\tau)}\simeq 5 \times 10^{11}$ GeV (evident from the intersection of $\langle\Gamma_\tau \rangle/{\mathcal{H}}$ line in blue with 1) while muon Yukawa interaction comes to equilibrium at $T^*_{0(\mu)} \simeq 10^9$ GeV as seen from $\langle\Gamma_\mu \rangle/{\mathcal{H}} =1$ point of Fig.~\ref{fig:1}. 

The lightest RHN with mass $M_1$ generates lepton asymmetry through leptogenesis in the type-I seesaw framework, described by the Lagrangian
\begin{equation}
	-\mathcal{L}= \overline{\ell}_{L_\alpha} (Y_{\nu})_{\alpha i} \tilde{H} N_{i}+ \frac{1}{2}  \overline{N_{i}^c}(M_{R})_{ii} N_i+ h.c.,
	\label{eq:1}
\end{equation}
where the lepton number violating Majorana mass term for RHNs, $M_R$, is considered to be diagonal, $M_R = {\rm{diag}}(M_1, M_2, M_3)$ for simplicity.  During the out-of-equilibrium decay process of the $N_1$, the charged lepton Yukawa interaction for one or more flavor(s) may enter equilibrium leading to the breaking of quantum coherence of the lepton doublet state along different flavor directions produced from the $N_1$ decay \cite{Barbieri:1999ma,Nardi:2006fx,Abada:2006fw,Blanchet:2006be,Dev:2017trv}. As a result, lepton asymmetry along individual flavors may start to become distinguishable. In this case, one needs to look for the evolution of the individual 
flavor lepton asymmetries ($Y_{\Delta_{\alpha = e, \mu, \tau}}$) instead of total lepton number asymmetry by constructing Boltzmann equations (BEs) for lepton asymmetries along individual flavors \cite{Nardi:2005hs,Nardi:2006fx}:

	\begin{align}
		s H z \frac{d Y_{N_1}}{dz}
		&=
		-
		\left( \frac{Y_{N_1}}{Y_{N_1}^{\rm eq}}	-	1	\right)
		\gamma_{D}  \,,\label{be1}
		\\
		s H z \frac{d Y_{\Delta_{\alpha}}}{dz}
		&=
		-\Bigg\{
		\left(	\frac{Y_{N_1}}{Y_{N_1}^{\rm eq}}	-1\right)
		\varepsilon_{\ell \alpha} \gamma_{D}
		+
		K^0_{\alpha} \sum_{\beta} \Bigg[ \frac{1}{2}  (C^{\ell}_{\alpha \beta} +
		C^H_{\beta}) \gamma_D \Bigg]\frac{Y_{\Delta_{\beta}}}{Y_{\ell}^{\rm eq}}
		\Bigg\},\label{eq:lep}
	\end{align}

\noindent where $K^0_{\alpha}= \frac{(Y_{\nu}^*)_{\alpha 1} (Y_{\nu})_{\alpha 1}}{(Y_{\nu}^{\dagger} Y_{\nu})_{11}}$ is known as flavor projector \cite{Nardi:2006fx,Blanchet:2006be} and $C^{\ell}, C^H$ matrices connect the asymmetries in lepton and Higgs sectors to asymmetries in $\Delta_{\alpha}$ expressed in terms of $Y_{\Delta{\alpha = e, \mu, \tau}}$ or $Y_{\Delta{\alpha = \kappa, \tau}}$ depending on the leptogenesis scale \cite{Nardi:2006fx}. Eq. \eqref{eq:lep} can be a set of two (three) equations if $10^9<M_1<5 \times 10^{11}$ GeV ($M_1< 10^9$ GeV). Here $\gamma_D$ represents the total decay rate density of $N_1$: 
\begin{align}
	\gamma_D= n_{N_1}^{\rm{eq}}\langle \Gamma_{N_1}\rangle, \qquad \langle\Gamma_{N_1} \rangle= \frac{K_1(z)}{K_2(z)}\frac{(Y_{\nu}^{\dagger} Y_{\nu})_{11}}{8\pi} M_1\,,
	\label{eq:17}
\end{align}
where $K_1(z)$ and $K_2(z)$ are the modified Bessels functions. $z=M_1/T$ is a dimensionless quantity, with respect to which we will look for the evolution of $Y_x$. The equilibrium number density of the $N_1$ can be expressed as:
\begin{align}
	n_{N_1}^{\rm{eq}}=\frac{g M_1^3}{2 \pi^2 z} K_2\left(z\right),
\end{align}
where $g$ is the number of degrees of freedom of $N_1$.


\section{Flavor effect in leptogenesis during reheating era}

Following inflation, the reheating discussion connects to the inflaton potential form near its minimum. We adopt a generic power-law form from \cite{Garcia:2020eof}, $V(\phi) = \lambda \frac{|\phi|^n}{M_P^{n-4}}$, originating from T-attractor models in no-scale supergravity \cite{Kallosh:2013hoa}. We focus on $n=2$, resembling Starobinsky inflation~\cite{Starobinsky:1980te,Ellis:2013xoa,Khalil:2018iip} in the large field limit. Here, without going into the detail of inflation (see \cite{Ellis:2015pla}), we just consider the parameter $\lambda \sim 2 \times 10^{-11}$ which is consistent with Planck+BICEP2/Keck (PBK) constraints on spectral index and tensor-to-scalar ratio at 95$\%$ C.L.~\cite{BICEP2:2018kqh}, yielding inflaton mass $m_{\phi} = (\partial ^2_{\phi} V(\phi))^{1/2} \simeq 1.5 \times 10^{13}$ GeV.

We introduce an effective inflaton-fermion interaction $y^f_{\phi} \phi \bar{f} f$ governing inflaton decay to radiation (massless SM fermions in the early Universe). This arises from dimension-5 operators involving SM Yukawa interactions: $\frac{\alpha_{\ell}}{\Lambda}\phi \bar{\ell}L H E_R, \quad \frac{\alpha_{u}}{\Lambda}\phi \bar{q}L \tilde{H} u_R, \quad \frac{\alpha_{d}}{\Lambda}\phi \bar{q}L H d_R$, where $\Lambda$ and $\alpha_{f}$ are the cut-off scale and coupling constants. This initiates inflaton decay to Higgs plus fermion-antifermion pairs, with subsequent fast Higgs decay (via SM Yukawa) effectively representing direct inflaton decay to fermion-antifermion pairs. Assuming all final state fermion-antifermion pairs ($g_f$ possibilities) couple uniformly to $\phi$, the total decay width is $\Gamma_{\phi} \simeq g_f^2 \frac{|y^f_{\phi}|^2}{8\pi} m_{\phi}$ (with $g_f = 30$ kept throughout).
We keep $y^f_{\phi}$ sufficiently small (quantified later) to exclude parametric excitation \cite{Greene:1998nh} and ensure perturbative $\phi$ decay. Reheating's role is thus limited to modifying the Universe's thermal history and its impact on charged lepton Yukawa ET.

The energy densities of the inflaton field ($\rho_{\phi}$) and radiation ($\rho_R$) satisfy the evolution equations\footnote{ Such scale factor dependence is explained in the Supplemental Material of \cite{Datta:2022jic}.}  in terms of $a$:
\begin{align}
	& \frac{d(\rho_\phi a^3)}{da}=- \frac{\Gamma_\phi}{\mathcal{H}}\rho_\phi a^2;~~{\rm{and}}~~ \frac{d\left(\rho_R a^4\right)}{da}= \frac{a^3}{\mathcal{H}} \Gamma_\phi \rho_\phi, 
	\label{rho-eq} 
\end{align}
where $\mathcal{H}^2 = ({\dot a}/a)^2 = (\rho_{\phi}+ \rho_{R})/3M^2_P$. At this moment, it is pertinent to define the temperature $T(a)$ connected to the $\rho_R$ by the relation
\begin{equation}
	T(a) = \left[ \frac{30}{g_*(a)\pi^2}\right]^{1/4} \rho^{1/4}_R(a).
\end{equation}
$T$ can therefore be estimated as a solution to the coupled Eqs. ~\eqref{rho-eq}. Note that at reheating onset, $\mathcal{H}$ is determined solely by $\rho_{\phi}$. Once the light decay products of $\phi$ thermalize, $\rho_R$ increases rapidly, reaching a maximum temperature $T_{\text{Max}}$. Subsequently, $\rho_R$ decreases gradually as $\rho_{\phi}$ diminishes through inflaton decay, until $\rho_R$ equals $\rho_{\phi}$ and begins dominating. This transition defines the reheating temperature $T_{\text{RH}}$.

We infer that when $T_{\rm{RH}} < T^{*}_{0(\alpha)}$ for flavor $\alpha$, equilibration (where $\langle \Gamma_{\alpha} \rangle = \mathcal{H}$) may occur within the reheating epoch spanning $T_{\text{Max}}$ to $T_{\text{RH}}$. The enhanced expansion rate during this period alters the $r.h.s$ of Eq.~\eqref{eq-temp} through ${\mathcal{H}}$, which now incorporates both $\rho_{\phi}$ and $\rho_{R}$ contributions. The modified $\langle \Gamma_{\alpha} \rangle/{\mathcal{H}}$ is plotted against $T$ in Fig. \ref{fig:1} (dotted lines) for $y^f_{\phi}=1.5\times 10^{-6}$. Prior to radiation domination, this displays distinct non-standard behavior (altered slope within $T_{\text{Max}}$ and $T_{\text{RH}}$) for each charged lepton flavor.

We note that $Y_{\tau}$ equilibrates at $T^*_{\tau} \sim 5 \times 10^{10}$ GeV, compared to $T^{*}_{0({\tau})} \simeq 5 \times 10^{11}$ GeV in the standard scenario, while $\mu_R$ and $e_R$ equilibrate at their standard temperatures $T^*_{0({\mu})}$ and $T^{*}_{0(e)}$. This occurs because $y_{\phi}^f = 1.5 \times 10^{-6}$ gives $T_{\text{RH}} = 1.86 \times 10^{10}$ GeV, which falls below $T^*_{0({\tau})}$, leaving $Y_{\mu}$ and $Y_{e}$ interactions unaffected by non-standard effects. For smaller $y_{\phi}^f$ values, $T_{\text{RH}}$ could drop below $T^*_{0({\mu})}$, yielding a modified $Y_{\mu}$ ET. Conversely, larger $y_{\phi}^f$ shortens the reheating period between $T_{\text{Max}}$ and $T_{\text{RH}}$, making any deviations in charged lepton ET less pronounced.

Having established this non-standard behavior during prolonged reheating, we now examine its implications for flavor leptogenesis.  With non-instantaneous reheating, three scenarios emerge: (a) $M_1 < T_{\text{RH}}$, (b) $T_{\text{RH}} < M_1 < T_{\text{Max}}$, and (c) $M_1 > T_{\text{Max}}$. Case (a) resembles standard thermal leptogenesis discussed in section~\ref{sec2}, while case (c), where $N_1$ cannot be thermally produced (we refrain from discussing it here and refer the readers to \cite{Barman:2021tgt}). We focus on case (b), where both thermal production and decay of $N_1$ occur during reheating.

Since $T_{\text{Max}} > M_1>T_{\rm RH}$, we must include $N_1$ energy density ($\rho_{N_1}$) in our analysis, as $N_1$ can be produced via inverse decay\footnote{For simplicity, we neglect scattering production.}. The modified Hubble parameter becomes
\begin{equation}
\mathcal{H}^2 = \frac{\rho_{\phi}+ \rho_{R} +\rho_{N_1}}{3M^2_P}.
\label{Hubble}
\end{equation}
Additionally, Eq.~\eqref{rho-eq} for $\rho_{R}$ acquires an extra term $\frac{a^3}{H} \langle\Gamma_{N_1}\rangle (\rho_{N_1}-\rho_{N_1}^{\text{eq}})$ on the $r.h.s.$, representing $\rho_R$ depletion from $N_1$ production via inverse decays. Here $\langle \Gamma_{N_1}\rangle$ is the $N_1$ decay rate \cite{Buchmuller:2004nz}, and $\rho_{N_1}$ evolves according to
\begin{equation}
\frac{d (\rho_{N_1} a^3)}{da}= -\frac{\langle\Gamma_{N_1}\rangle a^2}{\mathcal{H}}(\rho_{N_1}-\rho_{N_1}^{\text{eq}}).
\end{equation}

\noindent To solve the coupled equations for $\rho_{\phi}, \rho_R$, and $\rho_{N_1}$, we impose initial conditions $\rho_{N_1} = \rho_R = 0$ and set the post-inflationary energy density to $\rho_{\phi_{end}} = 3 V(\phi_{end})/2$, where $\phi_{end} = 0.78 M_P$ for this model class \cite{Garcia:2020eof}. We take $M_{2(3)}$ to be 100 times heavier than $M_{1(2)}$, precluding thermal production of $N_{2(3)}$.

For analysis, we evaluate the neutrino Yukawa $Y_{\nu}$ appearing in $\langle \Gamma_{N_1} \rangle$ with the help of Casas-Ibarra formalism \cite{Casas:2001sr}, 
\begin{align}
	Y_{\nu}=-i \frac{\sqrt{2}}{v} U D_{\sqrt{m}} \mathbf{R} D_{\sqrt{M}},
\end{align}
where $U$ is the PMNS \cite{Esfahani:2017dmu,Esteban:2020cvm,Zyla:2020zbs} mixing matrix diagonalizing $m_{\nu} = - Y_{\nu} M^{-1}_R Y^T_{\nu} v^2/2$ ($v =$ 246 GeV is the EW vev), $D_m (D_M) $ is the diagonal active neutrino (RHN) mass matrix and $\rm{\bf{R}}$ is a complex orthogonal matrix of the form as in \cite{Antusch:2011nz} parametrized by complex mixing angle $\theta_R$. The values of $\rm{Re}[\theta_R], \rm{Im}[\theta_R]$ are so chosen as to produce correct baryon asymmetry via leptogenesis and also to keep $Y_{\nu}$ entries perturbative. We incorporate the best fit values of mixing angles and mass-squared differences \cite{Esteban:2020cvm} to define $U$ and light neutrino mass eigenvalues (with $m_1 = 0$). 

Solving the coupled Boltzmann equations including $N_1$, we find that $\rho_{N_1}$ remains subdominant to $\rho_{R}$ (see \cite{Datta:2022jic}), due to its out-of-equilibrium $N_1$ decay in the same period. Consequently, $T_{\text{RH}}$ is essentially unaffected by this additional component. Recall from Fig. \ref{fig:1} that $Y_{\tau}$ equilibrates at $T^*_{\tau} = 5 \times 10^{10}$ GeV in this modified scenario. Thus, when $N_1$ decays around $T \lesssim M_1 = 10^{11}$ GeV, none of the charged lepton Yukawa couplings are in equilibrium, maintaining flavor coherence and producing unflavored leptogenesis. By contrast, standard thermal leptogenesis with the same $M_1$ occurs in a radiation-dominated era where $T^*_{0(\tau)} = 5 \times 10^{11}$ GeV $> M_1$, meaning $Y_\tau$ has already equilibrated and broken flavor coherence. In that scenario, lepton asymmetries would develop along two orthogonal directions: $\tau$ and $\kappa$ (a coherent superposition of $e$ and $\mu$ flavors) \cite{Barbieri:1999ma,Nardi:2005hs,Abada:2006fw,Nardi:2006fx,Blanchet:2006be}, yielding flavor leptogenesis. This transition from two-flavor to unflavored leptogenesis represents a significant result of our analysis.


Realizing that the case with $M_1 = 10^{11}$ GeV (and $y_{\phi}^f =1.5 \times 10^{-6}$) corresponds to unflavored leptogenesis scenario during this extended reheating period, we proceed to evaluate the $B-L$ asymmetry using the following Boltzmann equation \cite{Buchmuller:2004nz}, 
\begin{align}
	\frac{d (n_{\Delta} a^3)}{da}=-\frac{\langle\Gamma_{N_1}\rangle a^2}{\mathcal{H}}\left[\frac{\varepsilon_\ell}{M_1}(\rho_{N_1}-\rho_{N_1}^{\rm{eq}})+\frac{n_{N_1}^{\rm{eq}}}{2 n_\ell^{\rm{eq}}}n_{\Delta}\right],
\end{align}
with $n_{\Delta} = n_{B-L}$. However, in a more general case where a shift of regimes of thermal leptogenesis still leads to a flavored one, the corresponding equation would be

	\begin{align} 
		\frac{d(n_{\Delta_i}a^3)}{da}=-\frac{\langle\Gamma_{N_1}\rangle a^2}{\mathcal{H}}\left[\frac{\varepsilon_{\ell_i}}{M_1}(\rho_{N_1}-\rho_{N_1}^{\text{eq}})
		+\frac{1}{2}K^0_i\sum_j(C^\ell_{ij}+C^H_{j})\frac{n_{N_1}^{\rm{eq}}}{n_\ell^{\rm{eq}}}n_{\Delta_j}\right].
	\end{align}

 \noindent The CP asymmetry $\varepsilon_{\ell_{\alpha}}$ (for unflavored case, $\varepsilon_{\ell} = \sum_{\alpha} \varepsilon_{\ell_{\alpha}}$) involved is obtained from the decay of $N_1$ to a specific flavor $\ell_{\alpha}$ and estimated using standard expression~\cite{Covi:1996wh,Nardi:2006fx}. The final baryon asymmetry $Y_{B}$ is related to $n_{\Delta_i}$ by~\cite{Harvey:1990qw}:
$Y_{B}= \frac{28}{79}\sum_\alpha n_{\Delta_\alpha}/s.$ 

\begin{figure}[h]
	\centering
	{\includegraphics[width=0.5\linewidth]{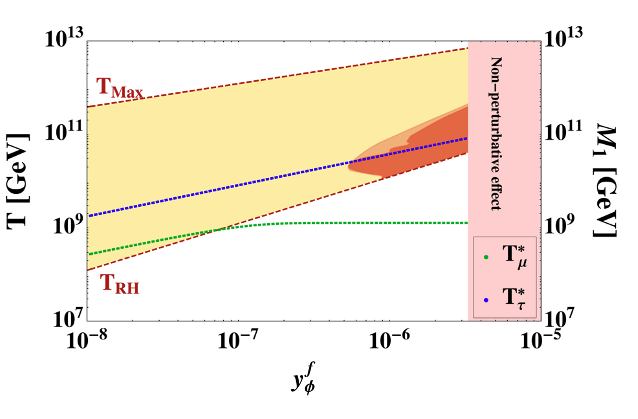}}
	\caption{Dark brown patch represents the allowed value of RHN mass $M_1$ and coupling $y_\phi^f$ which can produce the
correct baryon asymmetry for the scenario with modified flavor regimes. On the contrary, light brown patch
represents the allowed region when change in flavor effect was not taken into account. Here $y_\phi^f>3.3\times 10^{-5}$
(in light pink) indicates the nonperturbative regime}
	\label{fig:2}
\end{figure}

In Fig. \ref{fig:2}, we provide a scan in the $T-y_{\phi}^f$-$M_1$ plane  to examine different parameter choices within case (b), clarifying how the shifted ET affects thermal leptogenesis relative to the standard scenario. While we focus on effective inflaton-fermion interactions, inflaton-boson couplings such as $\sigma \phi^2 H^{\dagger}H$ and $\mu \phi H^{\dagger}H$ may also exist. However, to maintain perturbative inflaton decay and avoid rapid preheating particle production \cite{Kofman:1994rk}, these couplings must satisfy $\sigma < 5.5 \times 10^{-12}$ and $\mu < 5.2 \times 10^{-12}M_P$ \cite{Garcia:2020wiy}. With such constrained $\sigma$ and $\mu$ values, our analysis based solely on $y^f_{\phi} \phi \bar{f}f$ remains unaffected by inflaton-boson couplings.

\section{Conclusion}

We have demonstrated that prolonged reheating, spanning from the post-inflationary maximum temperature $T_{\text{Max}}$ to the reheating temperature $T_{\text{RH}}$, significantly delays the equilibration of charged lepton Yukawa interactions. The accelerated expansion during this period postpones the entry of these interactions into equilibrium, with the extent of this delay depending on the effective inflaton coupling to SM fields. Lower reheating temperatures (corresponding to weaker inflaton-SM couplings) extend the reheating era and substantially reduce the equilibration temperatures.

This phenomenon should apply to broader classes of SM interactions, opening avenues for further investigation. As one application, we examine its effect on leptogenesis where the lightest RHN in type-I seesaw is produced and decays out of equilibrium between $T_{\text{Max}}$ and $T_{\text{RH}}$. We find that delayed charged lepton Yukawa equilibration shifts the flavor regime of leptogenesis, a novel result warranting detailed study. This observation suggests several research directions. For instance, a follow-up study \cite{Datta:2023pav} identifies a new flavor leptogenesis regime in non-thermal leptogenesis during extended reheating. More broadly, we expect our findings to significantly impact low-scale leptogenesis scenarios with low reheating temperatures, with potential testability in current and upcoming experiments.	

\section{Acknowledgments}

I acknowledge financial support from the STFC Consolidated Grant ST/T001011/1. I also wish to thank Arunansu Sil and Arghyajit Datta for a very fruitful collaboration.

\bibliographystyle{JHEP}
\bibliography{ref,ref2}


\end{document}